\documentclass[twocolumn,twoside,slac_two]{revtex4}
\usepackage{graphicx}
\usepackage{fancyhdr}
\usepackage{amsmath}
\begin{document}

\title{Young radio sources: the duty-cycle of the radio emission and
  prospects for gamma-ray emission}

%


\author{M. Orienti, D. Dallacasa, G. Giovannini}
\affiliation{Dipartimento di Astronomia, Universit\`a di Bologna, via Ranzani
  1, 40127 Bologna, Italy}
\affiliation{INAF-IRA, via Gobetti 101, 40129 Bologna, Italy}
\author{M. Giroletti, F. D'Ammando}
\affiliation{INAF-IRA, via Gobetti 101, 40129 Bologna, Italy}

%

\begin{abstract}

The evolutionary stage of a powerful radio source originated
by an AGN is related to its linear size. In this context, compact
symmetric objects (CSOs), which are powerful and intrinsically small objects,
should represent the young stage in the individual radio source
life. However, the fraction of young radio sources in flux
density-limited samples is much higher than what expected from the
number counts of large radio sources. This indicates that 
a significant fraction of young radio sources does not develop to the classical
Fanaroff-Riley radio galaxies, suggesting an intermittent jet activity.
As the radio jets are expanding within the dense and
inhomogeneous interstellar medium, the ambient may play a role 
in the jet growth, for example slowing down or even disrupting 
its expansion when a jet-cloud interaction takes place. Moreover, 
this environment may provide the thermal seed 
photons that scattered by the lobes' electrons may be responsible for 
high energy emission, detectable by {\it Fermi}-LAT.

\end{abstract}

\maketitle

\thispagestyle{fancy}


\section{Introduction}
It is nowadays clear that powerful (L$_{\rm 1.4 GHz}$ $>$ 10$^{25}$
W/Hz) radio sources are a small fraction of the Active Galactic Nuclei
(AGN) generally associated with ellipticals, suggesting that the radio
activity is a transient phase in the life of these systems.
The onset of radio emission is
currently thought to be related to mergers which provide fuel to the
central AGN. The evolutionary stages of a powerful radio source are
related to its linear size. The discovery of the population of
intrinsically compact and powerful radio sources, known as compact
symmetric objects (CSOs), yielded to an
improvement of the models proposed to link the various evolutionary 
stages of the
radio emission. CSOs are characterized by linear sizes up to a
few kpc, and a synchrotron spectrum that turns over between hundreds
of MHz and the GHz regime. Their genuine youth has been proved by
estimate of both kinematic and radiative ages, which result to be
$\sim$10$^{3}$-10$^{4}$ years \citep{polatidis03,murgia03}. Their
radio morphology is dominated by mini-lobes/hotspots resembling a
scaled-down version of the classical edge-brightened
Fanaroff-Riley type-II galaxies \citep{fr74}. \\
Following the evolutionary models \citep[e.g.][]{fanti95}
CSOs should be the progenitors of the ``old'' FRII galaxies. However,
the excess of young objects in flux-limited samples suggests the
existence of short-lived objects unable to become FRII, and additional
ingredients, like the recurrence of the radio emission \citep{czerny09},
or the interplay between the source and the environment, must be
considered. Indeed, the dense and
inhomogeneous medium left by the merger that triggered the radio
emission may play a role in the source growth, for example
slowing down or even disrupting the jet expansion \citep{alexander00}. \\
Given their compact size, CSOs entirely reside within the innermost
region of the host galaxy. In the most compact radio
sources, the radio lobes are only a few parsecs from the AGN and their
relativistic electrons can scatter the thermal UV/IR 
seed photons produced by both the accretion disc and the torus, up
to high energies. For this reason, high energy $\gamma$-ray emission
detectable by {\it Fermi}-LAT
is expected from these compact objects.\\

\section{The duty-cycle of the radio emission}

\begin{figure*} [!thhh]
\begin{center}
\includegraphics[width=110mm]{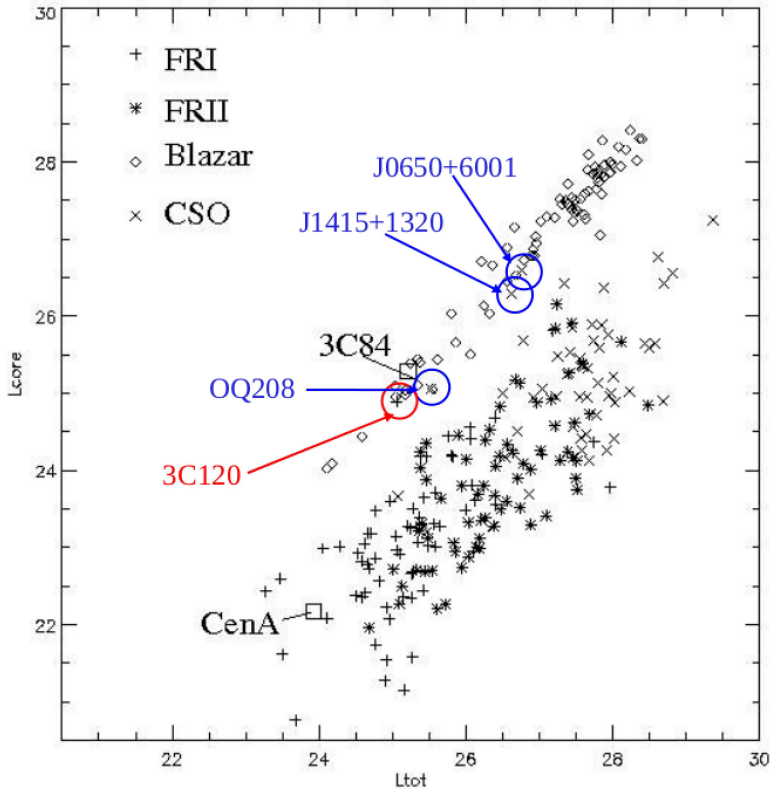}
\end{center}
\caption{Total luminosity ({\it x-axis}) vs core luminosity ({\it
    y-axis}) of a sample of FRI (+ signs) and FRII (asterisks) radio
  galaxies from \cite{gg01}, blazars
  (diamonds), and CSOs (x simbols). LBAS blazars are from \cite{gir10}.}
\label{luminosity}
\end{figure*}

When the sub-arcsecond morphology of compact symmetric objects could be
investigated by the advent of high spatial resolution observations, it
resulted to be characterized by the same structures typical of the
classical FRII galaxies, but on much smaller scale. For this reason,
\citep{mutel} suggested that CSOs should be objects whose radio
emission is still in a young phase of its evolution. Following this
approach, the fate of CSOs is to evolve into the FRII. However, the
number counts of young objects is too high with respect to those of
the ``old'' radio galaxies in flux-limited samples, even when a
luminosity evolution is taking into consideration. The discovery of
kpc-scale low-surface brightness structures, likely the fossil of an old episode
of radio emission, and connected with young radio
galaxies (i.e.~J0111+3906,\citep{baum90}) suggests that the radio
activity may be a recurrent phenomenon in the lifetime of a galaxy. 
Furtheremore, relics of previous activities have been recently found
also on pc-scales close to newly born objects indicating that the radio
activity may be also short-lived. This implies that the time elapsed
between two subsequent periods of radio activity can be as short as a
few thousand years (e.g.~J1511+0518, \cite{orienti08}), as
also predicted by the models (e.g. \cite{czerny09}). 
The radio source J1518+047
represents a clear example of a young (10$^{3}$ yr), but already
fading object \citep{orienti10}. In this radio source, only electrons
with $\gamma < 600$ are contributing to the radio emission, while
those with higher energies have already faded away. This indicates
that in young but fading objects no $\gamma$-ray emission is expected. \\
 
\section{The radio luminosity}

In terms of radio luminosity CSOs are comparable with the powerful
FRII galaxies. From Fig.~\ref{luminosity} it is clear that CSOs ({\it
  x symbols}) seem to extend at higher luminosity 
the correlation between the core
luminosity $L_{\rm core}$ and the total luminosity $L_{\rm tot}$ found
for FRI ({\it + signs}), and FRII ({\it
  asterisks}) by \citep{gg01}. Such high luminosities are expected
since CSOs are mainly found at higher redshift, between 0.4 and 2,
with only a few objects with $z \sim 0.1$. \\
A remarkable aspect pointed out in Fig.~\ref{luminosity} is the
presence of 3 CSOs, the radio galaxy OQ\,208 and the radio quasars
J0650+6001 and J1415+1320, in the region occupied by the sources, mainly
blazars, detected by {\it Fermi}-LAT during the first three-month observations
($diamonds$, \citep{abdo09}), making these CSOs good candidate for high
energy emission. It is worth noting that another misaligned object,
the FRI 3C\,120, is in the same LBAS region, and it was detected by
{\it Fermi}-LAT in 15-month observations \citep{abdo10}. \\ 

\subsection{The radio galaxy OQ\,208}

\begin{figure} [!hhh]
\begin{center}
\includegraphics[width=65mm]{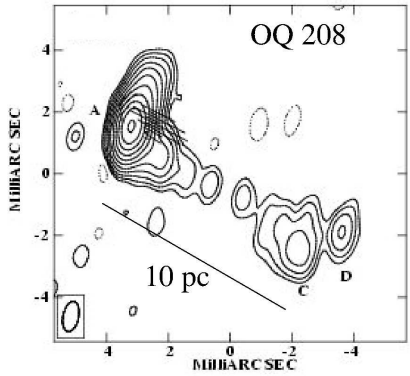}
\end{center}
\caption{VLBA images at 15 GHz of OQ208. Adapted from \cite{sta01}.}
\label{oq208_2}
\end{figure}

The radio source OQ\,208 is associated with a broad-line radio galaxy
at redshift $z$=0.076. Its radio luminosity is $\sim$2$\times$10$^{44}$ erg/s. It has an asymmetric triple radio structure 
of 10 pc in size and it is dominated by the western hotspot
(Fig. \ref{oq208_2}). 
A multi-epoch analysis of the changes of the pc-scale structure has
shown that the hotspots are separating with a velocity of
($0.2\pm0.1$)c, which provides a kinematic age for this source of
$\sim$160$\pm$60 yr.\\

\begin{figure} [!hhh]
\begin{center}
\includegraphics[width=60mm]{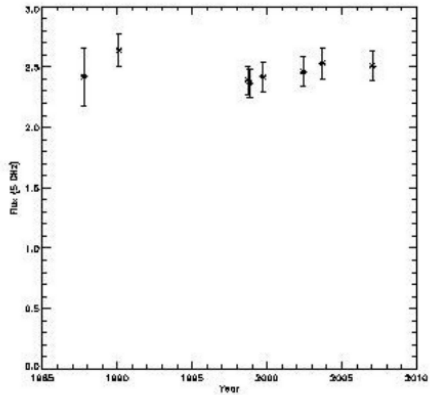}
\end{center}
\caption{The 5-GHz light curve of OQ\,208.}
\label{oq208}
\end{figure}

A low-surface brightness feature located about 40 mas from the main
structure is detected at low frequencies \citep{luo07}, and it marks
the fossil of a previous radio activity that took place a few thousand
years before the new episode \citep{orienti08} (Fig.~\ref{oq208_2}). The analysis of the 5 GHz
lightcurve from 1987 to 2007 does not show any significant flux
density variability (Fig.~\ref{oq208}). \\

\subsection{The radio quasar J0650+6001}

\begin{figure} [!bhhh]
\begin{center}
\includegraphics[width=60mm]{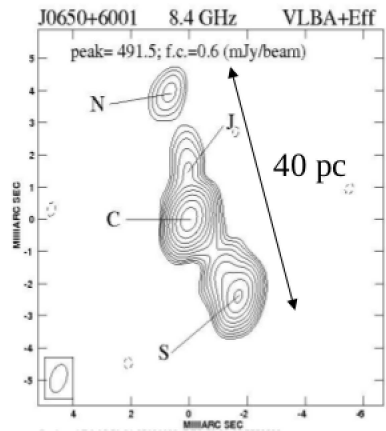}
\end{center}
\caption{VLBI image at 8.4 GHz of J0650+6001.}
\label{j0650}
\end{figure}

The radio source J0650+6001 is associated with a quasar at redshift
$z$=0.455. The source
is resolved into three components and its radio luminosity is $\sim$10$^{46}$ erg/s. The central component shows a flat spectrum, 
suggesting the presence of the core, while the two outer regions, 
with a steeper spectral index, display a highly asymmetric flux
density (Fig.~\ref{j0650}). Multi-epoch analysis of the changes in the pc-scale structure
shows that the outer components are separating with a velocity of
(0.39$\pm$0.19)c, which corresponds to a kinematic age of 360$\pm$170
years. Furthermore, the separation between the core component and the
southern hotspot seems to contract with an apparent velocity of
(0.37$\pm$0.02)c. Such contraction is interpreted in terms of a mildly
relativistic knot in the jet, still embedded in the central component, 
that is moving towards the southern component \citep{mo10}. 
This interpretation is
supported by the detection of some
variability related to the central component, as pointed up by the
5-GHz light curve \citep{mo10}.\\

\begin{figure} [!bhhh]
\begin{center}
\includegraphics[width=60mm]{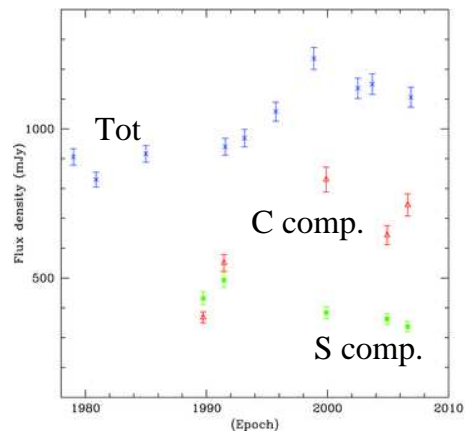}
\end{center}
\caption{The light curve of J0650+6001 at 5 GHz. Crosses indicate the source flux 
density from VLA data, while triangles and squares refer to VLBI flux 
density of component C and S, respectively. Adapted from \cite{mo10}.}
\label{j0650_2}
\end{figure}


\section{High energy emission in young radio sources}

Since CSOs completely reside within the host galaxy, a possible mechanism for
producing high energy emission may be the inverse Compton (IC) of thermal
UV/IR photons by the lobes' relativistic electrons \cite{st08}. Under the
assumption of equipartition and assuming that the jet luminosity is
L$_{\rm j}$ = 10$\times$L$_{\rm tot}$, 
and the luminosity provided by the UV photons is 
L$_{\rm UV}$ = 10$^{46}$ erg s$^{-1}$, 
we compute the expected luminosity at 1 GeV for OQ\,208 and J0650+6001 
using the formula from \citep{st08}:

\begin{equation} 
\begin{split} 
{[\varepsilon L_{\varepsilon}]_{\rm IC/UV} \over 10^{42} \, {\rm
    erg/s}} \sim 2 \, \left({\eta_{\rm e} \over \eta_{\rm B}}\right)
\left({L_{\rm j} \over 10^{45} \, {\rm erg/s}}\right)^{1/2} \left({LS
  \over 100 \, {\rm pc}}\right)^{-1} \times \\
\left({L_{\rm UV} \over 10^{46} \, {\rm erg/s}}\right) \left({\varepsilon \over 1 \, {\rm GeV}}\right)^{-0.25}
\label{lum}
\end{split}
\end{equation}
\noindent
or the appropriate IC/UV energy flux:
\begin{equation}
{[\varepsilon S_{\varepsilon}]_{\rm IC/UV} \over 10^{-12} \, {\rm erg / cm^{2} / s}} \sim 1.6 \times \left({[\varepsilon L_{\varepsilon}]_{\rm IC/UV} \over 10^{42} \, {\rm erg/s}}\right) \left({d_{\rm L} \over 100 \, {\rm Mpc}}\right)^{-2} ,
\label{flux}
\end{equation}
\noindent
where $d_{\rm L}$ is the luminosity distance to the source, LS is the source
linear size, $\eta_{\rm e}$/$\eta_{\rm B}$ is the ratio between the
particle energy and the magnetic field energy.\\
By means of Equations \ref{lum} and \ref{flux}, we can compute the
expected 1-GeV luminosity and flux density for OQ\,208 and J0650+6001,
which turn out to be: 

\begin{itemize}
\item OQ\,208: L = 2.8 $\times$ 10$^{44}$ erg s$^{-1}$, S = 3.8 $\times$
  10$^{-11}$ erg cm$^{-2}$ s$^{-1}$;

\item J0650+6001: L = 5 $\times$ 10$^{44}$ erg s$^{-1}$, S = 1.2 $\times$
  10$^{-12}$ erg cm$^{-2}$ s$^{-1}$ \\
\end{itemize}

\noindent Assuming standard parameters as above, OQ\,208, 
that is one of the closest CSOs, should
have been detected with the sensitivity obtained in one-year observations by
{\it Fermi}, i.e.~7$\times$10$^{-12}$ erg cm$^{-2}$ s$^{-1}$ at 5$\sigma$
computed assuming a photon index $\Gamma$=2.5. The non-detection of the source indicates that the parameters used in the model are too extreme, setting upper limits to the jet power and the amount of UV photons. 
The high redshift (z $>$ 0.4) typical of the majority of young radio sources makes these objects even more difficult to detect.
However, in the case of CSOs associated with steep-spectrum quasars, like
J0650+6001, where also moderate boosting effects should be present, we can
consider an additional contribution of IC made by relativistic electrons from
the jet (e.g.~\cite{gh05}). 

As {\it Fermi}-LAT continues to collect data and its sensitivity threshold improves, some young radio sources may be detected, giving us important information on the physical properties of these objects and the main mechanism responsible for their high energy emission.

%



\bigskip 
\begin{acknowledgments}

We acknowledge financial contribution from agreement ASI-INAF I/009/10/0.

\end{acknowledgments}

\bigskip 

\end{document}